\documentclass[pre,twocolumn,nofootinbib]{revtex4}
\usepackage{amsmath,amssymb}

\usepackage{graphicx}
\usepackage{dcolumn}
\usepackage{bm}
\usepackage{subeqnar}
\usepackage{array}
\usepackage{epstopdf,overpic,color,rotating}
\usepackage{caption}

\newcommand*{\horzbar}{\rule[.5ex]{2.5ex}{0.5pt}}
\graphicspath{{figures/}}

\begin{document}
\preprint{APS/123-QED}
\captionsetup[figure]{labelfont={bf},labelformat={default},labelsep=period,name={Figure}}
\captionsetup[table]{labelfont={bf},labelformat={default},labelsep=period,name={Table}}

\title{Discovering time-varying aeroelastic models of a long-span suspension bridge from field measurements by sparse identification of nonlinear dynamical systems}
\author{Shanwu Li$^1$, Eurika Kaiser$^2$, Shujin Laima$^1$, Hui Li$^1$,  Steven L. Brunton$^2$, and J. Nathan Kutz$^{3,}$\footnote{Electronic address: \texttt{kutz@uw.edu}}}%
\affiliation{$^1$ School of Civil Engineering, Harbin Institute of Technology, Harbin, China}
\affiliation{$^2$ Department of Mechanical Engineering, University of Washington, Seattle, WA. 98195-2420}
\affiliation{$^3$ Department of Applied Mathematics, University of Washington, Seattle, WA. 98195-2420}
\date{\today}

\begin{abstract}
We develop data-driven dynamical models of the nonlinear aeroelastic effects on a long-span suspension bridge from sparse, noisy sensor measurements which monitor the bridge.  Using the {\em sparse identification of nonlinear dynamics} (SINDy) algorithm, we are able to identify parsimonious, time-varying dynamical systems that capture vortex-induced vibration (VIV) events in the bridge.  Thus we are able to posit new, data-driven models highlighting the aeroelastic interaction of the bridge structure with VIV events.  The bridge dynamics are shown to have distinct, time-dependent modes of behavior, thus requiring parametric models to account for the diversity of dynamics.  Our method generates hitherto unknown bridge-wind interaction models that go beyond current theoretical and computational descriptions.  Our proposed method for real-time monitoring and model discovery allow us to move our model predictions beyond lab theory to practical engineering design, which has the potential 
to assess bad engineering configurations that are susceptible to deleterious bridge-wind interactions.  With the rise of real-time sensor networks on major bridges, our model discovery methods can enhance an engineers ability to assess the nonlinear aeroelastic interactions of the bridge with its wind environment.
\end{abstract}

\maketitle

\section{Introduction}\label{Introduction}

Through improved sensors and emerging network monitoring designs, it is now possible to continuously assess modern bridge performance in real time.  Not only is it critical that bridges be monitored, e.g. for traffic monitoring and safety, but the rich time series recordings provided by the sensors allow bridge engineers to gain new understanding of the nonlinear aeroelastic interactions of the bridge structure with wind disturbances.   The discovery of nonlinear dynamical systems from time series recordings of physical systems has the potential to revolutionize engineering efforts and provide new theoretical insights that are beyond the scope of current, state-of-the-art bridge models.  By leveraging sparse regression techniques, the so-called {\em sparse identification of nonlinear dynamics} (SINDy) method provides a new paradigm for data-driven model discovery~\cite{brunton_discovering_2016}.  The emergence of the SINDy algorithm is allowing researchers to discover governing evolution equations by sampling either the full or partial state space of a given system, respectively.   
Although nonlinear, data-driven system identification methods such as SINDy 
are emerging as viable techniques for a broad range of applications, the methods have yet to be applied to the complex aeroelastic interactions observed in bridges.  In this manuscript, we leverage (i) time-series measurements of a bridge sensor network, and (ii) the SINDy model discovery architecture to build data-driven models of the long-span suspension bridge.  We find that the SINDy architecture is effective in identifying parsimonious, time-varying dynamical systems which result from vortex-induced vibration (VIV) events in the bridge.  Thus we are able to posit new, data-driven models highlighting the aeroelastic interaction of the bridge structure with VIV events.

The conventional study of bridge aeroelastics is comprised of theoretical analysis, wind tunnel tests and computational fluid dynamics (CFD).  The complex fluid-structure interactions of bridges result in a variety of nonlinear, stochastic phenomena such as buffeting~\cite{davenport1962response,scanlan1978action,lin1979motion,lin1983multimode}, VIV~\cite{bishop1964lift,hartlen1970lift,iwan1974model,simiu1978wind}, and flutter~\cite{theodorsen1949general,bleich1950mathematical}. Despite tremendous advances in theoretical analysis and computational modeling, accurately characterizing bridge aeroelastics remains a challenging endeavor. Wind tunnel tests with cylinders, simplified sectional models or scaled, full aeroelastic models are combined with theoretical analysis to discover bridge aerodynamics \cite{nakamura1975unsteady,komatsu1980vortex,matsumoto1993mechanism,laima2013investigation,li2014reynolds}, leading to simplified semi-empirical models and a number of corresponding aerodynamic and aeroelastic parameter identifications \cite{scanlan1981state,davenport1992taut,scanlan1993problematics,sarkar1994identification,larose1998gust,chen2002advances,chen2000time,chen2001nonlinear,diana2008new,diana2010aerodynamic}. However, wind tunnel tests may suffer from uncertainties in the wind tunnels, such as uncertainties of equipment used to produce and measure wind, and results from different laboratories can differ even while using the same experimental models and under similar conditions \cite{sarkar2009comparative}. Significant progress has been made to replace these physical tests with computational models, resulting in a number of CFD-based methods \cite{larsen1998computer,ge2002investigation,bai2010three}.  However, not only are the computations exceptionally expensive, they rarely capture the quantitative dynamics correctly.  This is especially true for high Reynolds number flows and complex fluid-structure interactions that are typical for real bridge aerodynamics in the field.

Emerging data-driven methods are allowing for the discovery of physical and engineering principles directly from time-series recordings.  Our focus is on the SINDy architecture~\cite{brunton_discovering_2016}, which has been demonstrated on a diverse set of problems, including spatio-temporal~\cite{rudy2017data}, parametric~\cite{rudy2018data}, networked~\cite{mangan2016inferring}, control~\cite{brunton2016sparse}, and multiscale~\cite{champion2018discovery} systems.  Importantly, the SINDy architecture can be directly related to model selection theory~\cite{mangan2016inferring} in order to assess the quality and robustness of the model discovered.  The SINDy method is computationally efficient and the algorithms for all the innovations mentioned above are available as open source code.  An alternative data-driven approach to SINDy uses symbolic regression to identify directly the structure of a nonlinear dynamical system from data~\cite{bongard_automated_2007,schmidt_distilling_2009,cornforth_symbolic_2012}. This works remarkably well for discovering interpretable physical models, but the symbolic regression is computationally expensive and can be difficult to scale to large problems.  Deep neural networks (DNNs) are yet another approach to data-driven models, allowing for future-state prediction of dynamical systems~\cite{li_extended_2017,vlachas_data-driven_2018,yeung_learning_2017,Takeishi2017nips,Wehmeyer2017arxiv,Mardt2017arxiv,lusch_deep_2017,raissi2018multistep}.  However, a key limitation of DNNs, and similar data-driven methods, is the lack of interpretability of the resulting model:  they are focused on prediction and do not provide governing equations or clearly interpretable models in terms of the original variable set.

Our aim is to use the SINDy architecture to provide interpretable dynamical models that can aid in understanding the complex aerodynamic interactions of modern bridges, while providing critical insight for potential new bridge designs, models, and control.  Using bridge sensor network data, we discover the nonlinear dynamics that result from the complex interactions of VIV events with the bridge structure.  This fluid-structure interaction is beyond the description of current state-of-the-art bridge models and highlights the time-varying nature of the dynamics induced by wind variability.  In particular, we discover a parsimonious set of governing equations which are time-varying and driven by weak, moderate and strong gust disturbances.  Thus the magnitude of the VIV events changes the fundamental nature of the nonlinearity induced on the bridge structure.  Our models allow us to make future-state predictions and also identify distinct regimes of dynamical behavior.   Such models can help aid the design of new bridges as well as improve theoretical models of the nonlinear aeroelastic effects observed in real data.
 

The manuscript is outlined as follows:  In Sec.~II, VIV events are discussed in detail as they are the central concern affecting the nonlinear bridge aerodynamics.   Sec.~III details the bridge monitoring network and data acquisition of the time-series measurements used for model identification.  Sec.~IV develops the SINDy architecture for the bridge data of Sec.~III.  In this section, we discover new models that determine the bridge aerodynamics as a function of VIV events. In Sec.~V, we discover the distinct dynamical regimes of the bridge-wind system by clustering the discovered models. We also use these models in Sec.~VI to produce predictions and diagnostics for the entire VIV events.  The paper is concluded in Sec.~VII with an overview of the method and an assessment of the outlook of the method.

\section{Vortex Induced Vibration (VIV) of Long-Span Bridge}\label{VIV}
A long-span bridge may have intrinsically distinct modes of aeroelastic behavior such as buffeting, VIV and flutter. For modern bridges, flutter must be avoided in the design stage by increasing the critical flutter wind speed, because of its unique divergent response which results from aeroelastic instability. As a consequence, only buffeting and VIV are observed in modern bridges. Unlike buffeting, VIV involves aeroelastic effects characterized by fluid-structure interactions which result in a possible negative aerodynamic damping, thus generating large vibration amplitudes. VIV occurs during periodic vortex shedding within a range of shedding frequencies near the structural natural frequency. Large-amplitude oscillations occur in this range that appear to control the shedding process in a fluid-structure interaction phenomenon known as {\em lock-in}.

Comprehensive investigations of the mechanisms responsible for VIV have been performed. Nakamura and Mizota \cite{nakamura1975unsteady} have observed the {\em lock-in} phenomenon by measuring the lift force and characterizing wakes of rectangular prisms with various aspect ratios oscillating transversely in a uniform flow, with the short sides normal to the flow direction in a wind tunnel. It was found that the phase angles of the frequency response components of both the lift and near-wake velocity show abrupt changes when approaching the critical reduced wind velocity for vortex shedding. This is suggested to be a key phenomenon involved when solving the problem of the vortex excitation of bluff structures.  Komatsu and Kobayashi \cite{komatsu1980vortex} characterized two types of VIV through a series of experiments on various cross sections (such as L-shaped, T-shaped, H-shaped, and rectangular cylinders) with various aspect ratios in a wind tunnel. One is a forced small-amplitude vibration caused by von K\'arm\'an vortex shedding in cylinders (T-cylinders) with a separation point at the trailing edge. The other is a self-excited vibration with a relatively large amplitude in cylinders (L-, H- and rectangular cylinders) with a separation point at the leading edge, which occurs independent of the von K\'arm\'an vortex street. The generating mechanism in the latter case is described as a motion-induced vortex at the leading edge that synchronizes with the motion of the cylinder. The frequency of this type of vibration does not change within a certain range of wind velocities and coincides with the natural frequency of the cylinder, i.e. the {\em lock-in} phenomenon.  Li et al. \cite{li2014reynolds} have investigated the Reynolds number effects on the aerodynamic characteristics and VIV of a twin-box girder within a range of Reynolds number values ($ 5.85\times10^3 $ $\sim$ $ 1.12\times10^5 $). They find that the transition point of the separated shear layer moves upstream, and the bubble size gradually decreases with increasing Reynolds number values.  Such investigations give a strong foundation for a qualitative understanding of VIV and critical fluid-structure interactions.

In addition to understanding fundamental mechanisms, accurate VIV modeling is quite important, especially for the design of a bridge. Rigorous mathematical-physical modeling of VIV requires simultaneously solving the Navier-Stokes (N-S) equations and equations of motion of the structure. However, because of the strong nonlinearity of the N-S equations, this has proven mathematically and computationally intractable \cite{vickery1983across}. As a less-than-ideal alternative, simplified semi-empirical models have been proposed based on wind tunnel tests. To date, the most widely accepted empirical model is proposed by Simiu and Scanlan \cite{scanlan1981state}, which is described as
	\begin{align}\label{Eqn:SemiempiricalModel}
	m(\ddot{y}+2\zeta\omega_{1}\ddot{y}+\omega_{1}^{2}y)=F,
	\end{align}
	with
	\begin{align}
	F=&\frac{1}{2}{\rho}U^{2}(2D)\times\notag\\
&	\left(\underbrace{Y_{1}(K)\left(1-{\lambda}\frac{y^2}{D^2}\right)\frac{\dot{y}}{U}+Y_{2}(K)\frac{y}{D}}_{\text{motion induced}}+\underbrace{\frac{1}{2}\tilde{C}_{L}}_{\text{fluid induced}}\right)\label{Eqn:SemiempiricalModel:Force},
	\end{align}
where $ m $ is mass per unit span length, $ \omega_{1} $ is mechanical circular frequency, $ \zeta $ is mechanical damping ratio, $ y $ is cross-flow displacement, $ F $ is aerodynamic force, $ \rho $ is air density, $ U $ is wind speed, $ D $ is cross-flow dimension of the section, and $K={\omega}D/U $ is the reduced frequency of vortex shedding, where $ \omega $ is vortex-shedding frequency that satisfies the Strouhal relation, ${\omega}D/{U}=2{\pi}St$, outside lock-in regime and $St$ is the Strouhal number. The parameters $ \lambda, Y_{1}, Y_{2} $ and $ \tilde{C}_{L} $ have to be determined by calibration to experiments. Specifically, $ \lambda $ is a constant denoting the nonlinear dependence of self-excited force on displacement amplitude, $ \tilde{C}_{L} $ is the stochastic lift force coefficient, and $ Y_{1} $ and $ Y_{2} $ are aerodynamic derivatives.  The aerodynamic derivatives measure the change that occurs in an aerodynamic force acting on the structure when there is a small change in a parameter such as angle of attack, flow speed, etc.
 
The total force in the model consists of two types of forces: one is induced directly by vortex shedding around the bluff body simulated by the third term with $ \tilde{C}_{L} $ in Eq.~\eqref{Eqn:SemiempiricalModel:Force}, and the other is a motion-induced lift force represented by the first two terms in Eq.~\eqref{Eqn:SemiempiricalModel:Force} including aerodynamic damping with $ Y_{1} $ and aerodynamic stiffness with $ Y_{2} $. The direct forcing term with $ \tilde{C}_{L} $ is found to be small relative to the motion-induced force when large-amplitude oscillations are present~\cite{ehsan1990vortex}. The model~\eqref{Eqn:SemiempiricalModel} thus may be simplified by dropping the direct forcing term and then be normalized to:
\begin{eqnarray}\label{Eqn:SimplifiedModel}
\eta^{\prime\prime}(s)+2{\zeta}K_{1}\eta^{\prime}(s)+K^{2}_{1}\eta(s)&=&m_{r}Y_{1}\left[1-\lambda\eta^{2}(s)\right]\eta^{\prime}(s)\notag\\
&&+m_{r}Y_{2}\eta(s),
\end{eqnarray}
where $\eta=y/D$ is the normalized cross-flow displacement, $m_{r}={\rho}D^2/m$ is mass ratio, $K_{1}=\omega_{1}D/U$ is the reduced natural frequency, and primes indicate derivatives with respect to the dimensionless time, $s=Ut/D$.

A solution for the bridge dynamics is then sought in the form:
\begin{equation}\label{Eqn:Solution}
\eta(s)=A(s)\cos\left[Ks-\psi(s)\right].
\end{equation}
The VIV of a bridge is generally considered as {\em quasi-linear}, i.e., the system has a small amount of nonlinearity where $A(s)$ and $\psi(s)$ are slowly varying functions of dimensionless time $s$.  The solution ${\eta}(s)$ can then be replaced by two separate solutions for $A(s)$ and $\psi(s)$, which are given as follows:
\begin{subeqnarray}
\label{Eqn:ODEofAmplitude}
	&& A^{\prime}(s)=-\frac{1}{8}{\alpha}A(s)\left[A^{2}(s)-\beta^{2}\right],\\
\label{Eqn:ODEofPhase}
	&& \psi^{\prime}=\frac{1}{2K}\left[m_{r}Y_{2}+(K^{2}-K_{1}^{2})\right]s+\psi_{0},
\end{subeqnarray}
where $\alpha=m_{r}Y_{1}\lambda$, $\beta=({2}/{\sqrt{\lambda}})\left(1-({2{\zeta}K_{1}})/({m_{r}Y_{1}})\right)^{1/2}$
and $\psi_{0}$ is the initial phase.  This gives an asymptotic approximation for the VIV dynamics in the weakly nonlinear regime.  Unfortunately, it fails to hold for moderate and larger nonlinear interactions where our data-driven models are proposed to hold.

\section{Bridge sensor network:  Field measurements and data preprocessing}\label{Measurements}

The long-span suspension bridge investigated in this study crosses a narrow water channel that lies between two islands. A structural health monitoring system, including wind and vibration monitoring, was implemented in 2009 and has since continuously recorded measurements in real-time.

\begin{figure}
	\centering
	\includegraphics[width=1\linewidth]{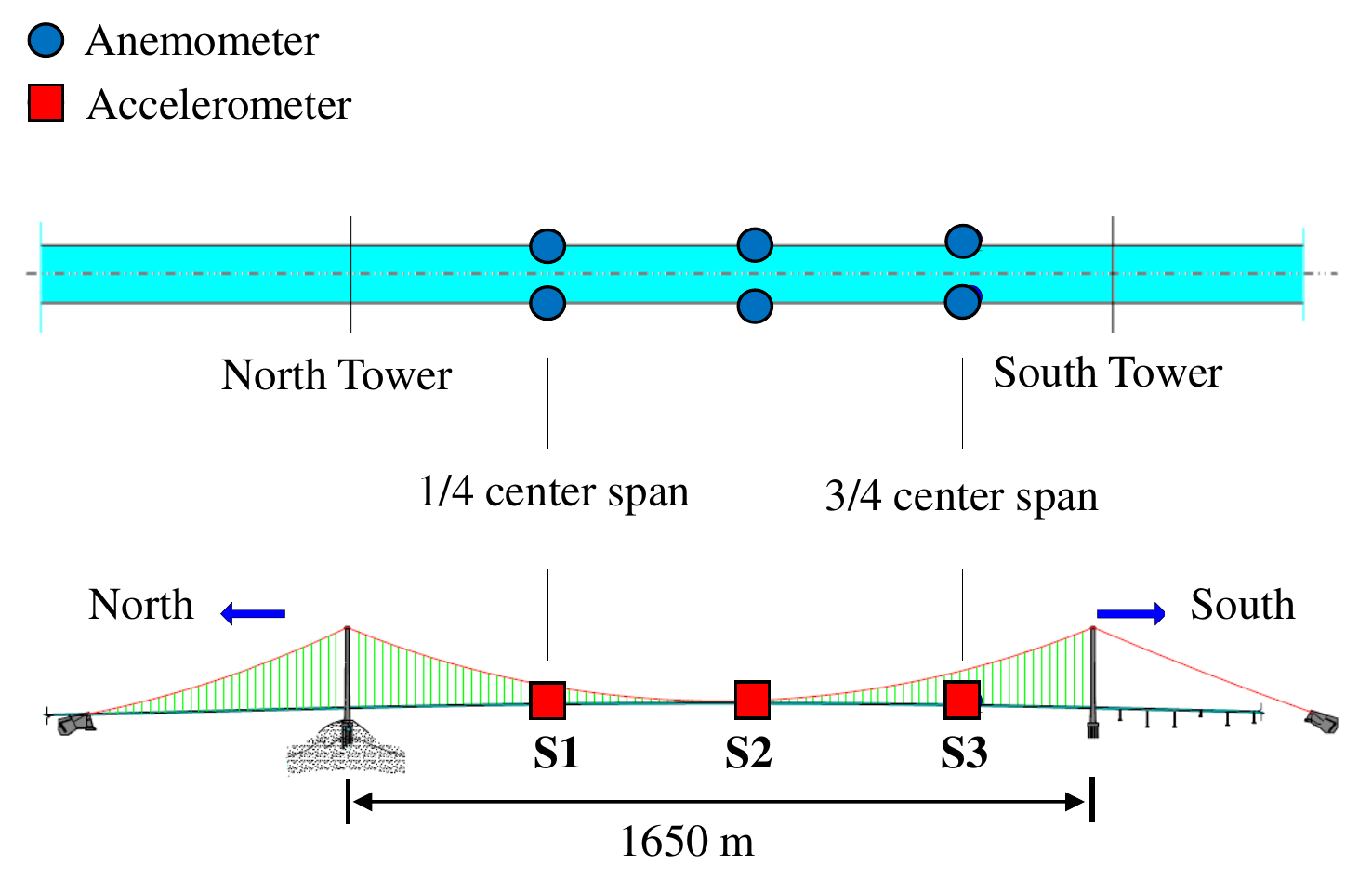}
	\caption{Sensor network on the bridge. Anemometers and accelerometers are installed at 1/4, 1/2 and 3/4 center span. Anemometers are installed on both sides of the bridge to measure the wind as it interacts with the bridge.}
	\label{Fig:SensorNetwork}
\end{figure}

At each side of the bridge, the wind speed and direction is monitored with anemometers.
In particular, Young Model 81000 three-dimensional ultrasonic anemometers with a sampling frequency of 32 Hz are located at 1/4, 1/2 and 3/4 center span (locations are indicated by S1, S2 and S3 in Fig.~\ref{Fig:SensorNetwork}, respectively) on both the upstream and downstream sides. These anemometers are installed on lighting columns at a height of 6 meter above the bridge deck surface. The wind data used in this
study are all from the inflow anemometers, which can measure natural winds without interference from bridge components. Vertical vibration of the bridge deck is monitored by GT02 force-balance triaxial accelerometers with a sampling frequency of 50 Hz at S1, S2 and S3.

\begin{figure}
	\centering
	\includegraphics[width=1\linewidth]{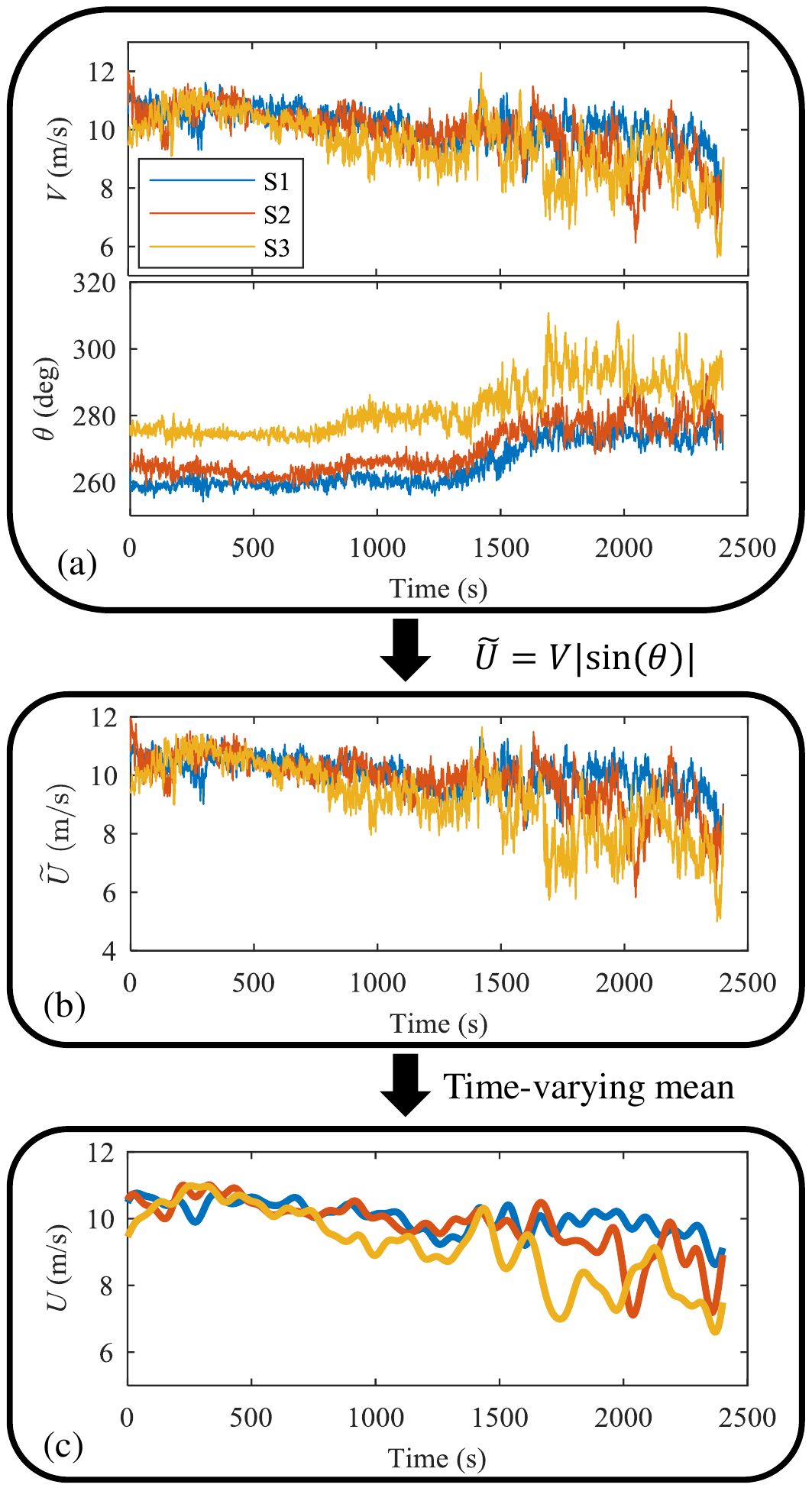}
	\caption{Preprocessing of wind data. (a) Horizontal instantaneous wind speed \textit{V} and wind direction $\theta$ are obtained from original measurements of wind speed. $90^{\circ}$ and $270^{\circ}$ indicate the perpendicular direction to the spanwise direction. (b) The wind speed component perpendicular to the spanwise direction is determined. (c) The time-varying mean wind speed is estimated by applying a low-pass filter.}
	\label{fig:wind}
\end{figure}

\begin{figure}
	\centering
	\includegraphics[width=1\linewidth]{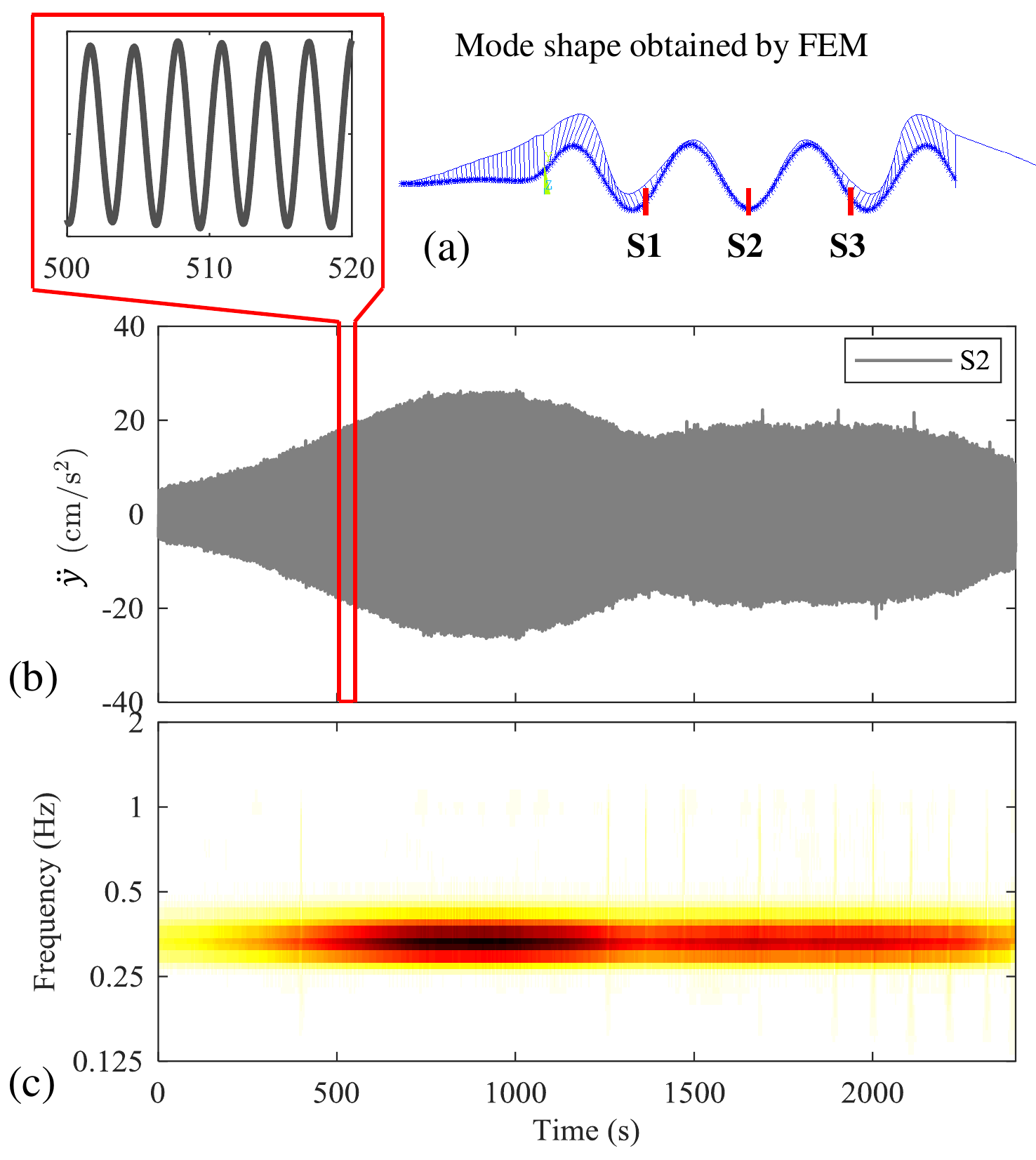}
	\caption{Time-frequency analysis of measured acceleration for a VIV event. (a) The mode shape of the bridge obtained by an accompanying numerical simulation using FEM for the measured mode of VIV with a modal frequency 0.327 Hz. (b) Acceleration history of a VIV event. (c) Time-frequency analysis of the vibration acceleration by the continuous wavelet transform.}
	\label{fig:acc}
\end{figure}

VIV events of this bridge captured by wind and vibration histories were identified using cluster analysis in a previous study~\cite{li2017cluster}. In the present study, we first process the original data to identify potential key factors accounting for the VIV aeroelastics. First, the wind data is pre-processed (see Fig.~\ref{fig:wind}). Histories of the horizontal wind speed $V$ and wind direction $\theta$ are obtained from the measured horizontal wind components. The component perpendicular to the spanwise direction is obtained by $\tilde{U}=V\left|{\sin}(\theta)\right|$.  The time-varying mean wind speed $U$ is determined by applying a low-pass filter to $\tilde{U}$. Further, we analyze the time-frequency characteristics of vibration (see Fig.~\ref{fig:acc}). It is found that the amplitude shifts with time, while the frequency stays almost constant during the entire event. This result indicates that $\left[Ks-\psi(s)\right]$ in Eq.~\eqref{Eqn:Solution} does not lead to a frequency shift for a real VIV; and we thus only need to focus on the time-varying amplitude $A(s)$ in Eq.~\eqref{Eqn:Solution}. Accordingly, the ordinary differential equation (ODE) of $A$ described by Eq.~\eqref{Eqn:ODEofAmplitude} is the key equation describing the VIV aeroelastics and is also the most important prior knowledge for data-driven modeling in this paper.

\begin{figure}
	\centering
	\includegraphics[width=1\linewidth]{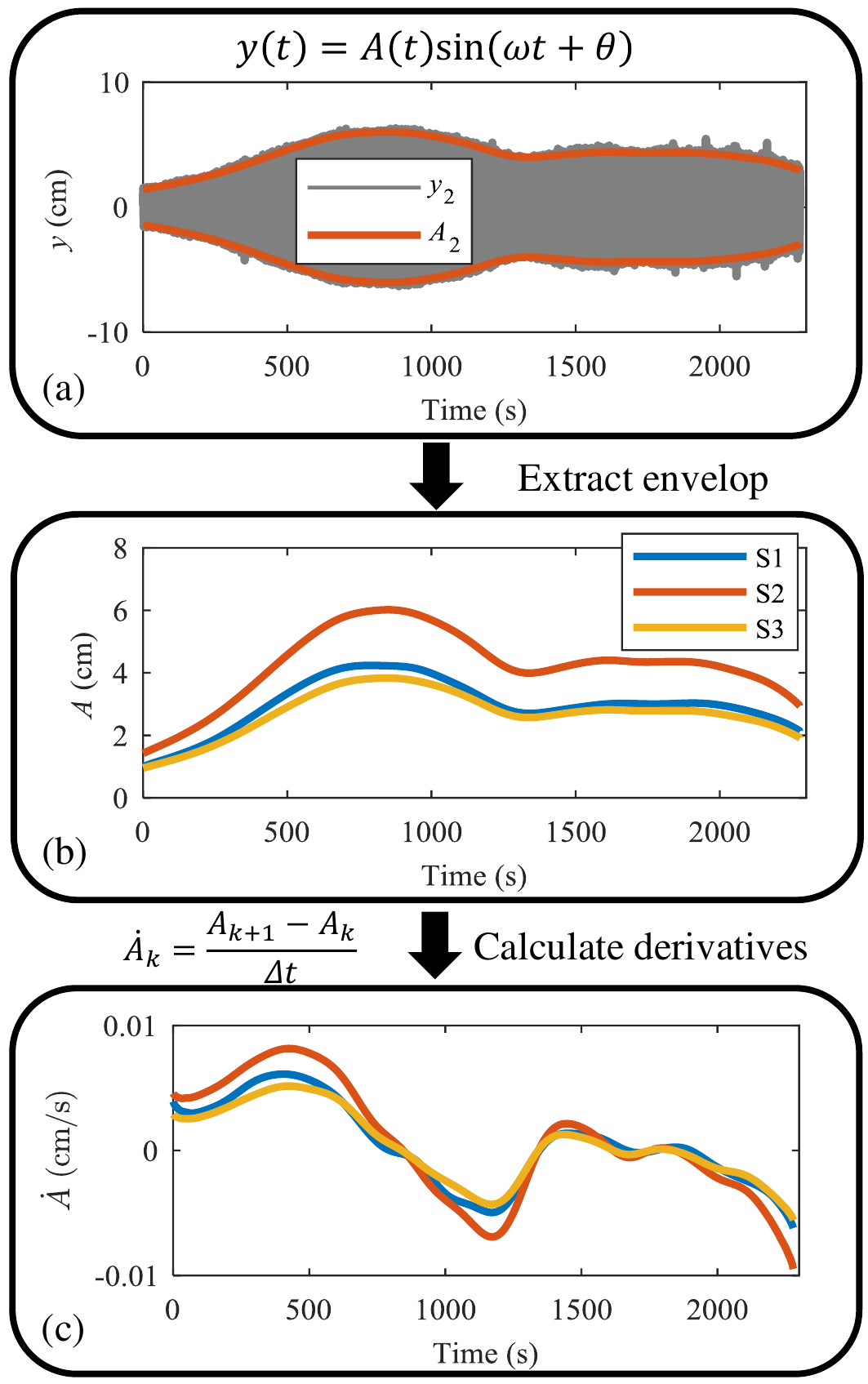}
	\caption{Preprocessing of vibration data. (a) The time-varying displacement amplitude \textit{A} is obtained by extracting the envelop from the displacement history \textit{y} which is obtained by integration of acceleration $ \ddot{y} $ in the frequency domain. (b) Vibration amplitudes are obtained for all the three sensor locations. (c) Time derivatives of the amplitudes are obtained.}
	\label{fig:dis}
\end{figure}

However, Eq.~\eqref{Eqn:ODEofAmplitude} needs to be generalized from wind tunnel tests to field measurements by carefully considering three key points: (1) the wind condition during an entire VIV event is time-varying for real VIVs while constant during wind tunnel tests; (2) the spatial dimension of the aeroelastic system for field measurements, which depends on the constellation of the sensors, is higher than the one-dimensional section model typically used in wind tunnel tests, and (3) the real bridge produces strongly nonlinear responses, unlike the weakly nonlinear theory. 
To build our data-driven model and account for these considerations, we extract the envelop of the vibration displacement to obtain the time-varying displacement amplitude $A$ and its time derivative $\dot{A}$ (see Fig.~\ref{fig:dis}).

\section{Data-driven model discovery:  Sparse identification of time-varying, nonlinear dynamics}\label{SINDy} 
 
We use data-driven model discovery methods to extract improved characterizations of the nonlinear aeroelastic bridge-wind interactions.  Our aim is to make maximal use of the time-series data generated by the bridge sensors.
 
\subsection{The SINDy Algorithm}

The primary method used for our model discovery is the SINDy algorithm, which leverages advances in machine learning and sparse regression to discover nonlinear dynamical systems from data~\cite{brunton_discovering_2016}.  
SINDy solves an overdetermined linear system of equations by sparsity-promoting regularization.  The basic algorithmic structure of SINDy has been modified to discover parametrically-dependent systems~\cite{rudy2018data}, resolve multiscale physics~\cite{champion2018discovery}, infer biological networks~\cite{mangan2016inferring}, discover spatio-temporal systems~\cite{rudy2017data}, and identify nonlinear systems with control~\cite{brunton2016sparse,Kaiser2017arxivB}.

Consider a dynamical system of the form
\begin{eqnarray}
\dot{\mathbf{x}} = \boldsymbol{f}(\mathbf{x})\label{eq:f}
\end{eqnarray}
where the function $\boldsymbol{f}(\cdot)$ is unknown, but assumed to have only a few dominant contributing terms.  The 
SINDy algorithm posits a large set of potential candidate functions that comprise $f(\cdot)$, then uses a sparsity-promoting regression to determine the dominant terms.   The relevant active terms in the dynamics can be solved for using an $\ell_1$-regularized regression that penalizes the number of active terms.  The general framework for SINDy is shown in Fig.~\ref{fig:time-varying SINDy}(b).

Sensor measurements are used to collect time-series data which  are arranged in the data matrix:
\begin{eqnarray}
\mathbf{X} = \begin{bmatrix} \mathbf{x}(t_1) & \mathbf{x}(t_2) & \cdots & \mathbf{x}(t_m)\end{bmatrix}^T, \label{eq:xmat}
\end{eqnarray} 
where the superscript `$T$' denotes the matrix transpose. 
The matrix $\mathbf{X}$ is $m\times n$, where $n$ is the dimension of the state $\mathbf{x}\in\mathbb{R}^n$ and $m$ is the number of measurements of the state in time.  
Similarly, the matrix of derivatives
\begin{eqnarray}
\dot{\mathbf{X}} = \begin{bmatrix} \dot{\mathbf{x}}(t_1) &  \dot{\mathbf{x}}(t_2) & \cdots &  \dot{\mathbf{x}}(t_m)\end{bmatrix}^T,
\end{eqnarray} 
is collected or computed from the state data in $\mathbf{X}$.  Accurate derivatives are critical for model identification, and the total-variation regularized derivative~\cite{chartrand_numerical_2011} is used as a numerically robust method to compute derivatives from noisy data.  

A library of candidate nonlinear functions is constructed from $\mathbf{X}$.  This takes the general form
\begin{eqnarray}
\boldsymbol{\Theta}(\mathbf{X}) = \begin{bmatrix} \mathbf{1} & \mathbf{X} & \mathbf{X}^2 & \cdots & \mathbf{X}^d  & \cdots &   \sin(\mathbf{X}) & \cdots  \end{bmatrix},\label{Eq:NLLibrary}
\end{eqnarray}
where $\mathbf{X}^d$ denotes the matrix containing all possible column vectors obtained from time-series of the $d$-th degree polynomials in the state vector $\mathbf{x}$.  For example, for a system with two states $\mathbf{x} = \begin{bmatrix} x_1, & x_2\end{bmatrix}^T$, the quadratic terms are given by the matrix $\mathbf{X}^2 = \begin{bmatrix} x_1^2(\mathbf{t}), & (x_1x_2)(\mathbf{t}), & x_2^2(\mathbf{t})\end{bmatrix}$, where $\mathbf{t}$ is a vector of times at which the state is measured.  Thus, the vector $\mathbf{x}$ is a symbolic \emph{variable}, while the matrix $\mathbf{X}$ is a \emph{data} matrix.

It is now possible to relate the time derivatives in $\dot{\mathbf{X}}$ to the candidate nonlinearities in $\boldsymbol{\Theta}(\mathbf{X})$ by:
\begin{eqnarray}
\dot{\mathbf{X}} = \boldsymbol{\Theta}(\mathbf{X})\boldsymbol{\Xi},\label{Eq:SINDy1}
\end{eqnarray}
where each column $\boldsymbol{\xi}_k$ in $\boldsymbol{\Xi}$ is a vector of coefficients that determines which terms are active in the $k$-th row in Eq.~(\ref{eq:f}).  Sparsity promoting algorithms are used to ensure that most of the entries of the column $\boldsymbol{\xi}_k$ are zero.  SINDy promotes sparsity by sequential least-squares thresholding, which has recently been shown to converge under suitable conditions~\cite{Zheng2018arxiv,zhang2018convergence}.  

By identifying the sparse coefficient vectors $\boldsymbol{\xi}_k$, a model of the nonlinear dynamics may be constructed:
\begin{eqnarray}
\dot{x}_k = \boldsymbol{\Theta}(\mathbf{x})\boldsymbol{\xi}_k,
\end{eqnarray}
where $x_k$ is the $k$th element of $\mathbf{x}$ and $\boldsymbol{\Theta}(\mathbf{x})$ refers to a row vector whose elements are symbolic functions of $\mathbf{x}$, as opposed to the data matrix $\boldsymbol{\Theta}(\mathbf{X})$.

Using sparse regression to identify active terms in the dynamics from the candidate library $\boldsymbol{\Theta}(\mathbf{X})$ is a convex optimization.  
The alternative is to apply a separate constrained regression on every possible subset of nonlinearities, and then to choose the model that is both accurate and sparse.   
This brute-force search is intractable, and the SINDy method makes it possible to select the sparse model in this combinatorially large set of candidate models.

\subsection{Time-Varying SINDy}

The potential for the SINDy algorithm to discover nonlinear dynamics has been demonstrated on a diverse set of problems~\cite{mangan2016inferring, brunton2016sparse,champion2018discovery}. However, the dynamics in these problems are often assumed to not change with time, i.e.\ they generally have constant coefficients, although the original SINDy algorithm is able to account explicitly for forcing and parameterized dynamics. More recently, SINDy has been extended to deal with parametric partial differential equations~\cite{rudy2018data} by allowing the coefficients $\boldsymbol{\xi}$ of each term in the library to be time-dependent. In the present study, we propose a time-varying SINDy to discover intrinsically and strongly time-varying dynamics: 
\begin{eqnarray}
\dot{\mathbf{x}} = \boldsymbol{f}_t(\mathbf{x})\label{eq:f_timevarying}
\end{eqnarray}
where $\boldsymbol{f}_t$ changes with time, but is not assumed as an explicitly time-dependent function.
Then, the coefficients of the terms identified with SINDy are time-varying so that the active terms can vary dramatically with time:
\begin{eqnarray}
\dot{x}_k = \boldsymbol{\Theta}(\mathbf{x})\boldsymbol{\xi}_k(t).
\end{eqnarray}
We assume $t\in[t-w,t]$ with window size $w$ over which the coefficient vector $\boldsymbol{\xi}_k(t)$ is determined.
The basic idea is shown in Fig.~\ref{fig:time-varying SINDy}. We introduce a time sampling window $w$ which moves across the time series data collected from a time-varying dynamical system (see Fig.\ref{fig:time-varying SINDy}(a)), and conduct a basic SINDy regression on the data in the window at each time step (see Fig.\ref{fig:time-varying SINDy}(b)).  We can then sort the obtained active terms and corresponding coefficients in order to reveal the intrinsically time-varying dynamics.

\subsection{SINDy to Model Aeroelastic Bridge Effects}

The time-independent model of VIV described by Eq.~\eqref{Eqn:SimplifiedModel} only accounts for a simple laboratory experiment where the wind speed is within the {\em lock-in} range.  This would give time-independent constants for the SINDy parameters. However, the real VIV of a prototype bridge in the field is typically a time-varying, nonlinear dynamical system characterized by the time-varying aerodynamic regime which results from the variability of natural wind forcing. Equation~\eqref{Eqn:SimplifiedModel} thus fails in simulating real VIV events. In the present study, we propose a time-varying SINDy model to discover the time-varying bridge aerodynamics from measured VIV events of a bridge. As mentioned in Section~\ref{Measurements}, the response of the VIV shows no frequency shift during an entire event (see Fig.~\ref{fig:acc}), indicating that only the time-varying amplitude must be modeled to represent the VIV response. 

\begin{figure*}
	\centering
	\includegraphics[width=.9\linewidth]{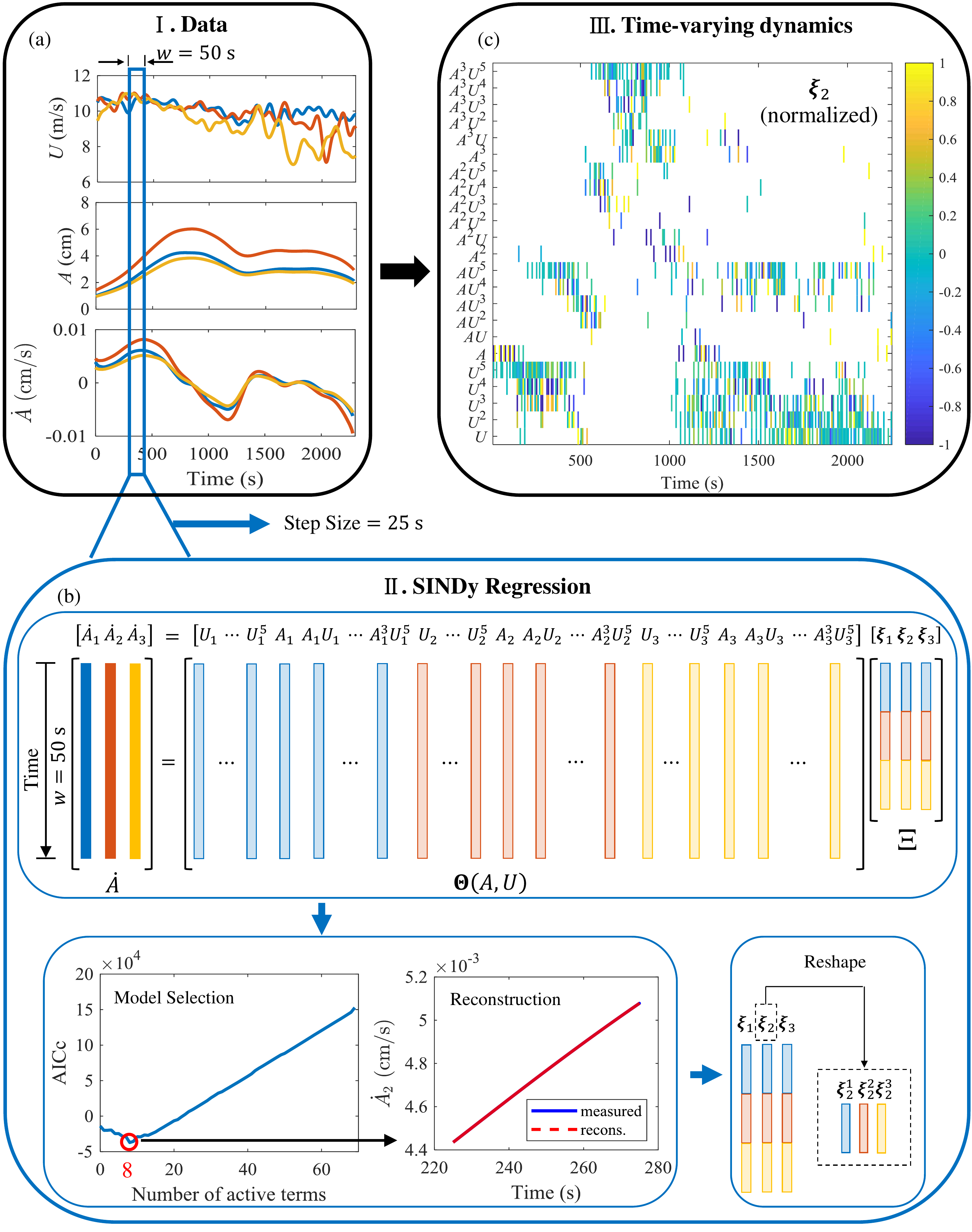}
	\vspace{-.1in}
	\caption{Schematic of the time-varying SINDy framework, demonstrated on the aerodynamics of a VIV event on a bridge. (a) Data is collected from the measurement system, including a history of time-varying mean wind speed \textit{U}, amplitudes \textit{A} and time derivatives $ \dot{A}$. (b) A typical SINDy is conducted in a moving time window at each time instant. The time window is swept across the entire VIV event with a size of 50 seconds and a moving step size of 25 seconds. Each component of the obtained model $ \boldsymbol{\xi} $ is reshaped into a 3-column  matrix, where each column corresponds to sensor measurements at one bridge section, respectively, for a more interpretable representation of the obtained time-varying aerodynamics. (c) A time series of the model in terms of $\boldsymbol{\xi}$ is obtained that captures the time-varying aerodynamics of an entire VIV event.}
	\label{fig:time-varying SINDy}
\end{figure*}

The input to the time-varying SINDy algorithm consists of time-series data of time-varying mean wind speeds ${\bf U}$, vibration displacement amplitudes of the bridge deck ${\bf A}$, and the time derivatives $\dot{\bf A}$ obtained by numerical differentiation for a measured VIV event. 
Here, the wind speed $U_k$ and displacement amplitude $A_k$, $k=1,2,3$, denote the respective measurement at the $k$th sensor location along the bridge. In particular, the subscripts 1, 2 and 3 indicate the sensor locations at the bridge sections S1, S2 and S3, respectively.
We learn the time-parametrized model over a short-term window with a duration of 50 seconds, which moves across the VIV event time-line with a step size of 25 seconds, as shown in Fig.~\ref{fig:time-varying SINDy}(a). 
A SINDy regression is then performed for data in each 50 second time window, as shown in Fig.~\ref{fig:time-varying SINDy}(b).  Although the analytic model in Eq.~\eqref{Eqn:ODEofAmplitude} is unable to predict the time-varying dynamics of a entire VIV event, it guides our construction of candidate functions for the library $\boldsymbol{\Theta}$.  Specifically, we expand the terms in Eq.~\eqref{Eqn:ODEofAmplitude} and propose a set of polynomial products of the time-varying mean wind speed ${\bf U}$ and the vibration displacement amplitude ${\bf A}$:
\begin{equation}\label{Eqn:Terms}
{\bf A}^{i}\odot{\bf U}^{j},
\end{equation}
where $i=0,1,2,3$, $j=0,1,2,3,4,5$ are the element-wise power, and $i$ and $j$ do not both equal zero.

In addition to the choice of polynomial terms, the single-section model characterizing wind tunnel tests with Eq.~\eqref{Eqn:ODEofAmplitude} is generalized to a higher-dimension variant by incorporating the sensors placed at the S1, S2 and S3 along the bridge span.  After computing time derivative data $\dot{{\bf A}}$, the proposed SINDy architecture takes the form  
\begin{equation}\label{Eqn:SINDyonBridge}
\dot{\bf A}(t) = \boldsymbol{\Theta}({\bf A}, {\bf U})\, \boldsymbol{\Xi}(t),
\end{equation}
where the library of candidate functions is defined by
\begin{equation}
\boldsymbol{\Theta}^T({\bf A},{\bf U}) = \begin{bmatrix}
\horzbar & {\bf A}& \horzbar\\
\horzbar &{\bf U}& \horzbar\\
\horzbar & {\bf A}^2& \horzbar\\
\horzbar &{\bf A}\odot{\bf U}& \horzbar\\
\horzbar & {\bf U}^2& \horzbar\\
\horzbar &{\bf A}^2\odot{\bf U}&\horzbar\\
 &\vdots& \\
\horzbar &{\bf A}^3\odot{\bf U}^5 &\horzbar
\end{bmatrix}   
\end{equation}
and `$\odot$' denotes the element-wise multiplication ofc ${\bf A}$ and ${\bf U}$, e.g.\ ${\bf A}\odot{\bf U} = [A_1U_1,\,A_2U_2,\,A_3U_3]$.
Note that the dynamics of $A_k$ at the $k$th location depend on sensor information at all three locations, i.e. they depend on $A_l$ and $U_l$ with $l=1,2,3$.

The {\em Akaike information criterion} with a correction for small sample sizes (AICc)~\cite{cavanaugh1997unifying} is proprosed to aid the SINDy regression for model selection. 
Trajectory reconstructions are then used to ensure the accuracy of the model. 
For interpretability and visualization, we reshape the obtained models $\boldsymbol{\xi}_k$ into three sets of
models $\boldsymbol{\xi}_k^{l}$, $l=1,2,3$ corresponding to sensors at locations S1, S2 and S3.  For example, $\boldsymbol{\xi}_k^{\rm{1}}$ is a vector of coefficients of terms $\left[U_{1}, \dots, A^{3}_{1}U^{5}_{1}\right]$.

\begin{figure}
	\centering
	\includegraphics[width=1\linewidth]{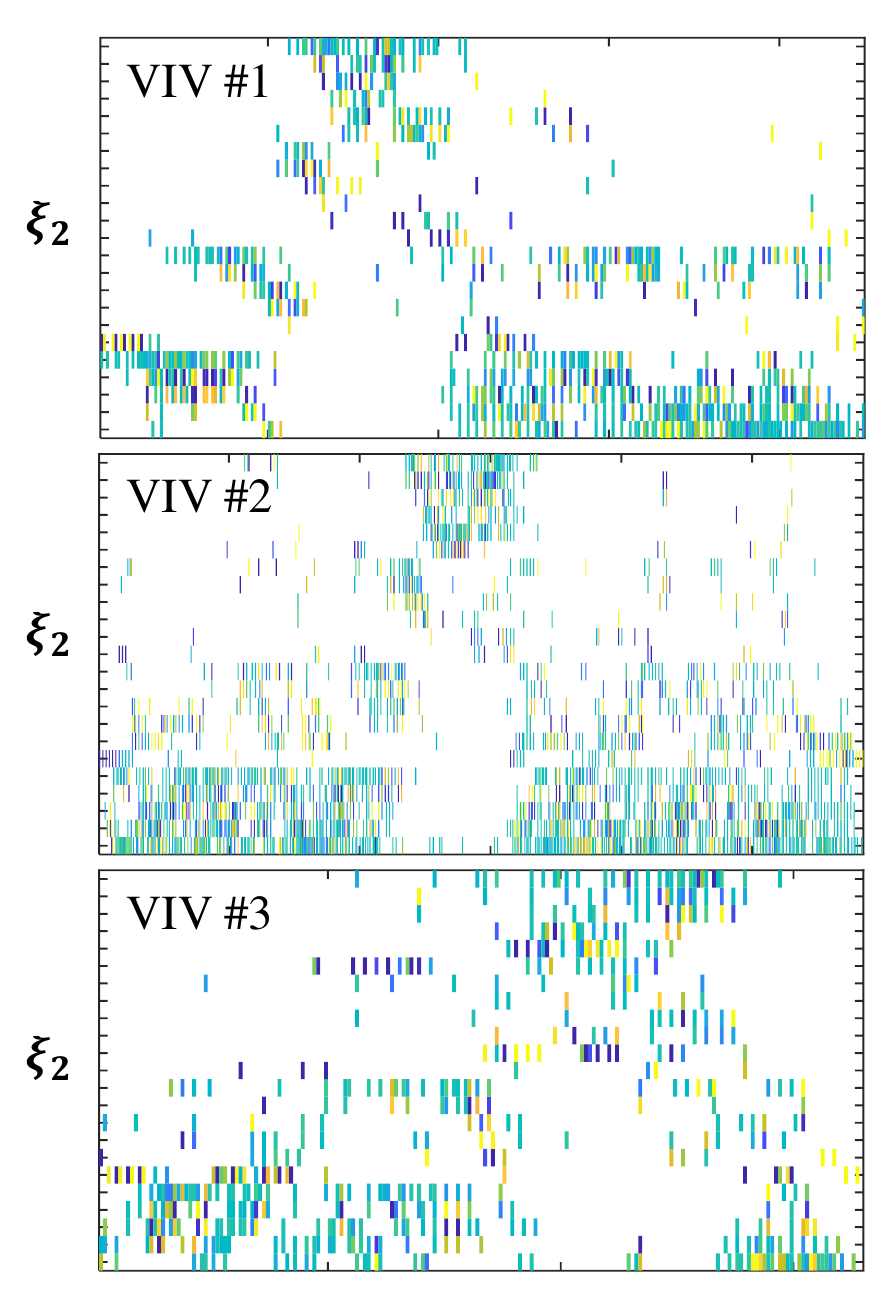}
	\caption{Time-varying dynamics of three exemplary VIV events discovered by time-varying SINDy.}
	\label{fig:3 VIV events}
\end{figure}

SINDy results in a set of models for a VIV event after the 50 second time window moves through the entire event, as shown in Fig.~\ref{fig:time-varying SINDy}(c). It is found that the active terms and coefficients vary significantly with time.  It should be noted that the candidate terms are sorted in an ascending polynomial order of vibration amplitude $A$ and wind speed $U$ from bottom to top in Fig.~\ref{fig:time-varying SINDy}(c), and that a higher polynomial order of $A$ implies a stronger (more nonlinear) wind-structure interaction with more motion-induced (self-excited) effects. In the same way, we have conducted the proposed time-varying SINDy on 31 measured VIV events in total and report the results for three VIVs in Fig.~\ref{fig:3 VIV events}. From the obtained time-varying dynamics for all the VIV events, We can intuitively find 4 dynamical regimes which are distinguished by the polynomial order of vibration displacement amplitudes $\bf A$. Accordingly, we rewrite the time-varying SINDy model (see~ Eq.~\eqref{Eqn:SINDyonBridge}) for these discovered different dynamical regimes specifically and respectively in Table~\ref{SINDyModelforRegime}. The dynamics at any moment during a VIV event must be from one of or the mix of the discovered regimes.

\begin{table*}
	\centering
	\renewcommand\arraystretch{1.5}
	\begin{tabular}{p{3cm} p{10cm} p{4cm}}
		\hline
		Dynamical Regime & SINDy Model  & Characteristics \\
		\hline
		Regime 1 & $\dot{\bf A}(t) = \left[{\bf U}, {\bf U}^2, {\bf U}^3, {\bf U}^4, {\bf U}^5, {\bf A}\right] \boldsymbol{\Xi}_{R1}(t)$ & No self-excited effect.\\
		\hline
		Regime 2 & $\dot{\bf A}(t) = \left[{\bf A}\odot{\bf U}, {\bf A}\odot{\bf U}^2, {\bf A}\odot{\bf U}^3, {\bf A}\odot{\bf U}^4, {\bf A}\odot{\bf U}^5, {\bf A}^2\right] \boldsymbol{\Xi}_{R2}(t)$ & Slight self-excited effect.\\
		\hline
		Regime 3 & $\dot{\bf A}(t) = \left[{\bf A}^2\odot{\bf U}, {\bf A}^2\odot{\bf U}^2, {\bf A}^2\odot{\bf U}^3, {\bf A}^2\odot{\bf U}^4, {\bf A}^2\odot{\bf U}^5, {\bf A}^3\right] \boldsymbol{\Xi}_{R3}(t)$ & Medium self-excited effect.\\
		\hline
		Regime 4 & $\dot{\bf A}(t) = \left[{\bf A}^3\odot{\bf U}, {\bf A}^3\odot{\bf U}^2, {\bf A}^3\odot{\bf U}^3, {\bf A}^3\odot{\bf U}^4, {\bf A}^3\odot{\bf U}^5\right] \boldsymbol{\Xi}_{R4}(t)$ & Strong self-excited effect.\\
		\hline	
	\end{tabular}
	\caption{The obtained SINDy models for the discovered different dynamical regimes which are distinguished by the polynomial order of vibration displacement amplitude $\bf A$ in the active terms accounting for the level of self-excited effect in the bridge-wind interaction. $\boldsymbol{\Xi}_{R1}(t) $, $ \boldsymbol{\Xi}_{R2}(t) $, $ \boldsymbol{\Xi}_{R3}(t) $ and $ \boldsymbol{\Xi}_{R4}(t) $ are the corresponding subsets of $ \boldsymbol{\Xi}(t) $, respectively.}
	\label{SINDyModelforRegime}	
\end{table*}

\begin{figure}
	\centering
	\includegraphics[width=1\linewidth]{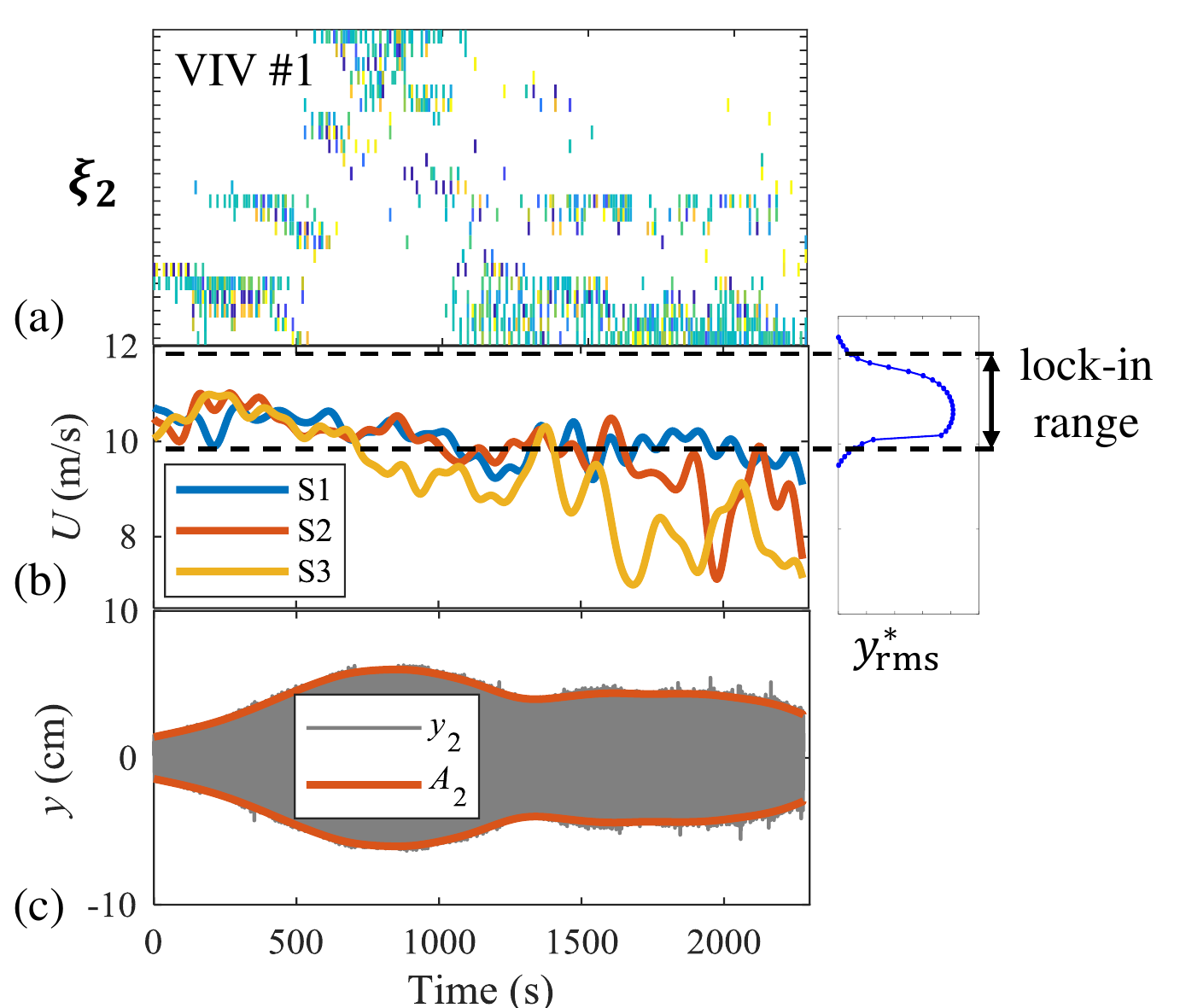}
	\caption{Interpretation of time-varying aerodynamics for a VIV event and the "lock-in" range obtained in the~\cite{li2017cluster}. (a) The time series of model sets $\boldsymbol{\xi}_{2}$. (b) The history of time-varying mean wind speeds \textit{U} compared with the "lock-in range". (c) The history of displacement $y_2$ with the amplitude $A_2$.}
	\label{fig:interpretation}
\end{figure}

In the {\em lock-in} range of VIV for the bridge considered in~\cite{li2018data}, we find a strong correlation between the time variation of aerodynamics and wind speed (see Fig.~\ref{fig:interpretation}). Specifically, during the first stage (0 s $\sim$ 600 s), the wind speeds at S1, S2 and S3 all stay within the {\em lock-in} range, resulting in a full development of wind-structure interaction with an increasing motion-induced (self-excited) effect.  This is indicated by the increasing polynomial order of $A$ in active terms with time.  During the second stage (600 s $ \sim$ 1000 s), the aerodynamic system reaches the steady state of high wind-structure interaction with the strong motion-induced (self-excited) effects.   Here, only the terms with the highest polynomial order in $A$ are active.  During the  third stage (1000 s $\sim$ end), wind speeds at S2 and S3 fall out of the {\em lock-in} range, resulting in a significant decrease of motion-induced (self-excited) effects. This is indicated by the decreasing polynomial order of $A$ in the active terms, i.e. the system becomes weakly nonlinear.   The obtained time-dependent, nonlinear dynamics is capable of producing a parsimonious model of the aerodynamics of a real VIV bridge event.

\section{Distinguished Dynamical Regimes:  Clustering of dynamic models}

The analysis of the identified time-varying aeroelastic responses from the VIV events (see Fig.~\ref{fig:interpretation} for a single VIV event) indicates the existence of several distinct dynamical regimes, all of which contribute to revealing the underlying, time-varying aeroelastic physics. 
The patterns associated with different model structures, e.g. as shown in Table.~\ref{SINDyModelforRegime}, indicate distinct dynamical regimes. This motivates the application of cluster analysis on the different model sets to automatically discover these different modes of dynamical aeroelastic behavior in the VIVs.

\subsection{Clustering algorithm}

In the clustering algorithm \cite{rodriguez2014clustering} applied in this study, two quantities are calculated for each data point $i$: the local density $\rho_{i}$ and the distance $\delta_{i}$. The local density of data point $i$ is defined as
\begin{subequations}\label{Eqn:Local Density}
	\begin{align}
	\rho_{i}&=\sum_{j}{\chi}\underbrace{(d_{ij}-d_{c})}_{=:D}\\
	{\chi}(x)&=\left\{
	\begin{array}{lr}
	1{\quad}if{\quad}D<0\\
	0{\quad}otherwise
	\end{array}
	\right.,
	\end{align}
\end{subequations}
where $d_{ij}$ is the distance between data point $i$ and $j$, $d_{c}$ is a cutoff distance, and $\chi(\cdot)$ is a step function. The quantity $\rho_{i}$ measures the number of points that are closer than $d_{c}$ to data point $i$. The distance $\delta_{i}$ is defined as the minimum distance between the point $i$ and any other point with a higher density:
\begin{equation}\label{Eqn:Distance}
\delta_{i}=\mathop{\min}_{j:\rho_{j}>\rho_{i}}(d_{ij}).
\end{equation}
But for the point with the highest global density, the distance $\delta_{i}$ is defined as the maximum distance between data point $i$ and any other point as there is no data point with a higher density.

By plotting all the data points with the two quantities defined by Eq.~\eqref{Eqn:Local Density} and Eq.~\eqref{Eqn:Distance}, the cluster centers are recognized fast and easily as points for which the value of $\delta_{i}$ is anomalously large without a definite pre-specified number of clusters. 
After the identification of cluster centers, each remaining point is assigned to the same cluster as its nearest neighbor of higher density. It is noted that this algorithm is sensitive only to the relative magnitude of $\rho$ for different points and the clustering results are robust against the parameter $d_{c}$ for large datasets~\cite{rodriguez2014clustering}.


\subsection{Cluster analysis of dynamic models of bridge aerodynamics}

\begin{figure}
	\centering
	\includegraphics[width=0.9\linewidth]{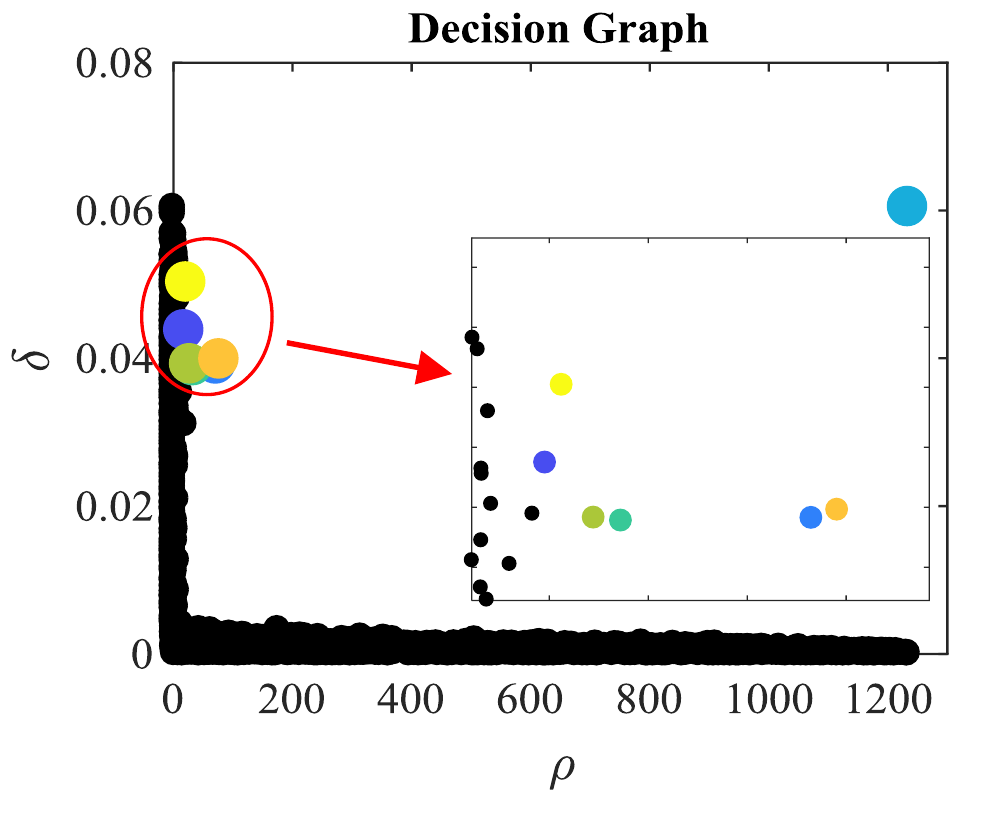}
	\caption{Seven cluster centers in the model sets determined based on the Decision Graph.}
	\label{fig:decision graph}
\end{figure}

We consider the time series of the model coefficient vector $\boldsymbol{\xi}_2$ obtained by the time-varying SINDy algorithm for each of the measured 31 VIV events. Each model set (consisting of $\xi_{2}^{1}$, $\xi_{2}^{2}$, and $\xi_{2}^{3}$) is considered as a data point in the 23-dimensional model space for this cluster analysis, where each dimension corresponds to a term in the candidate function library. With ploting all the model sets with the two quantities defined by Eq.~\eqref{Eqn:Local Density} and Eq.~\eqref{Eqn:Distance}, seven cluster centers are identified as points for which the value of $\delta_{i}$ is anomalously large (see Fig.~\ref{fig:decision graph}). And the correponding clusters are obtained after the assignment of each remaining model set to the same cluster as its nearest neighbor of higher density.

\begin{figure*}
	\centering
	\includegraphics[width=1\linewidth]{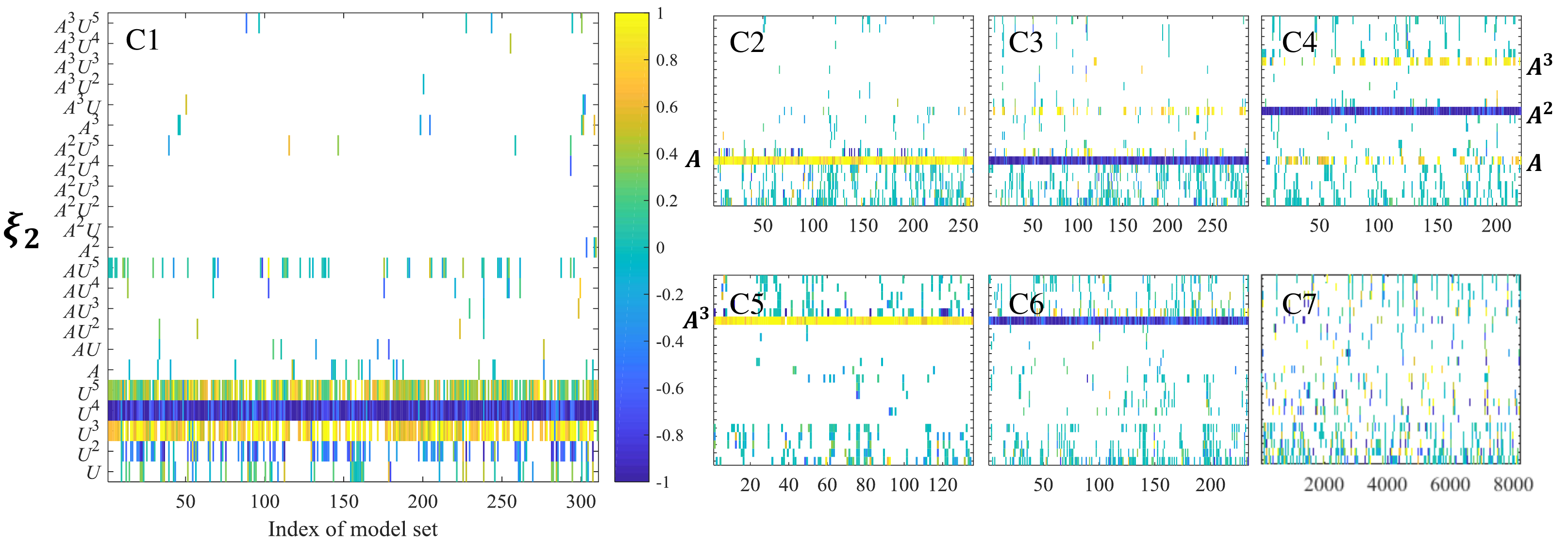}
	\caption{The obtained clusters in the model sets.}
	\label{fig:clusters}
\end{figure*}

\begin{figure*}
	\centering
	\includegraphics[width=1\linewidth]{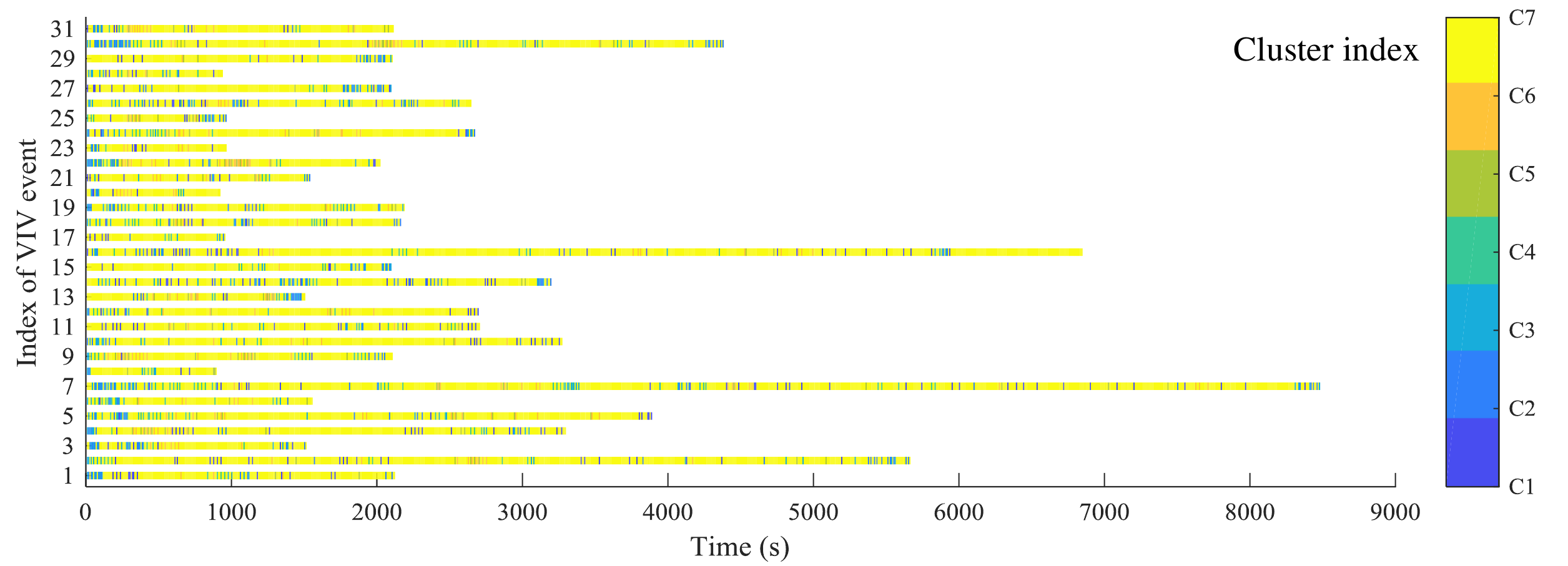}
	\caption{Representation of time-varying aerodynamics for all the VIV events by the obtained clusters.}
	\label{fig:representation by clusters}
\end{figure*}

The obtained clusters along with their members are shown in Fig.\ref{fig:clusters}. It can be found that model sets in the same cluster have common dominant terms. Specifically, the common dominant terms in C1 are $U$, $U^2$, $U^3$, $U^4$ and $U^5$, indicating pure forced vibrations by wind. C2 and C3 are dominated by the same term $A$, however with different signs. Thus, these clusters represent linear dynamics with respect to $A$. The most dominant term in C4 is $A^2$, followed by $A$ and $A^3$. C5 and C6 have the same dominant term $A^3$, which is correlated with the parameter $\alpha$ in Eq.~\eqref{Eqn:ODEofAmplitude} involving the aerodynamic derivative $Y_{1}$ in Eq.~\eqref{Eqn:SemiempiricalModel:Force}. We can thus know that the discovered terms with polynomial of $A^3$ are actually consistent with the aerodynamic damping component of the motion induced force in the Simiu and Scanlan's model (See Eq.~\eqref{Eqn:SemiempiricalModel:Force}). In C7 no term is dominant, but instead the dynamics are mixed where both wind-induced force and self-excited force account for the vibration of the bridge. It can be found that these clusters are distinguished by the polynomial order of the vibration amplitude $A$ in the dominant terms, which is just consistent with the intuitively discovered 4 dynamical regimes shown in Table~\ref{SINDyModelforRegime}, indicating that different dynamical regimes in VIV aerodynamics of this bridge are distinguished by the level of motion-induced (self-excited) effect in the wind-structure interaction. As analyzed with Fig.~\ref{fig:interpretation}, the temporal dynamical regime of the bridge-wind system is determined by the temporal wind condition and bridge state.
Based on the clustering result, the time-varying aerodynamics for all the VIV events can be represented in a simplified way with a cluster index (see Fig.~\ref{fig:representation by clusters}).

\section{Simualation and Forecasting for the VIV of the Long-Span Bridge}

\begin{figure*}
	\centering
	\includegraphics[width=1\linewidth]{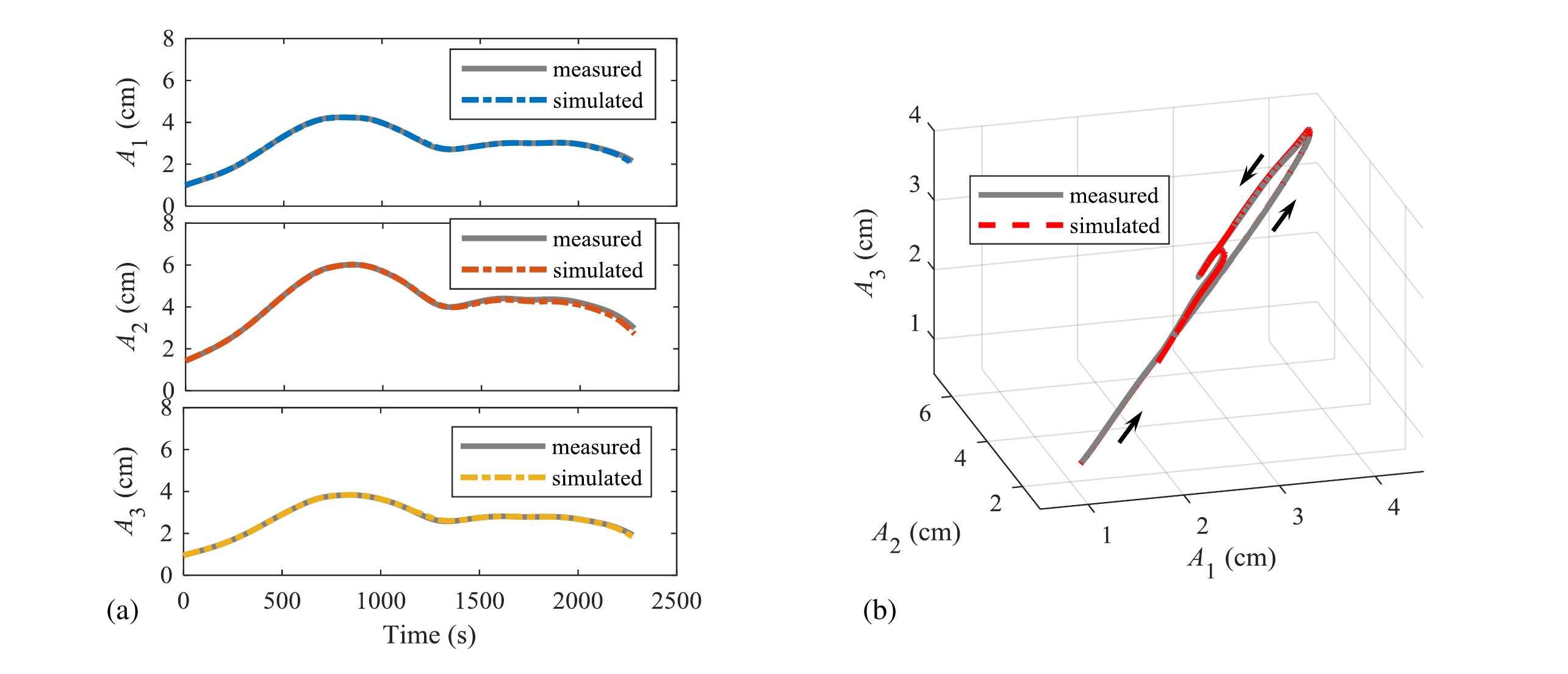}
	\caption{Simulation for the entire VIV event No. 1 by solving the ODE with the obtained time-dependent parameters ${\bf{\Xi}}(t)$ (see Eq.~\eqref{Eqn:SINDyonBridge}) with only the measured initial state ${\bf A}(t=0)$ and the measured wind history ${\bf U}(t)$ given. (a) Time histories of spatial components of ${\bf A}$. (b) Trajectory of ${\bf A}$ in the phase space.}
	\label{fig:simulation}
\end{figure*}

We have obtained a specific parametric model for each measured VIV event by the proposed time-varying SINDy. Each VIV is thus represented by an ODE with the corresponding time-dependent parameter $ \boldsymbol{\Xi}(t) $ (see Eq.~\eqref{Eqn:SINDyonBridge}).  To validate the obtained model, we simulate the entire VIV event by numerically solving the parametric model with the corresponding time-dependent parameter $ \boldsymbol{\Xi}(t) $ given the measured initial state ${\bf A}(t=0)$ and the measured wind history ${\bf U}(t)$. The comparison between the simulated and measured states show a near perfect agreement (see Fig.~\ref{fig:simulation}), indicating the correct modeling of the nonlinear time-varying dynamics.

\begin{figure*}
	\centering
	\includegraphics[width=1\linewidth]{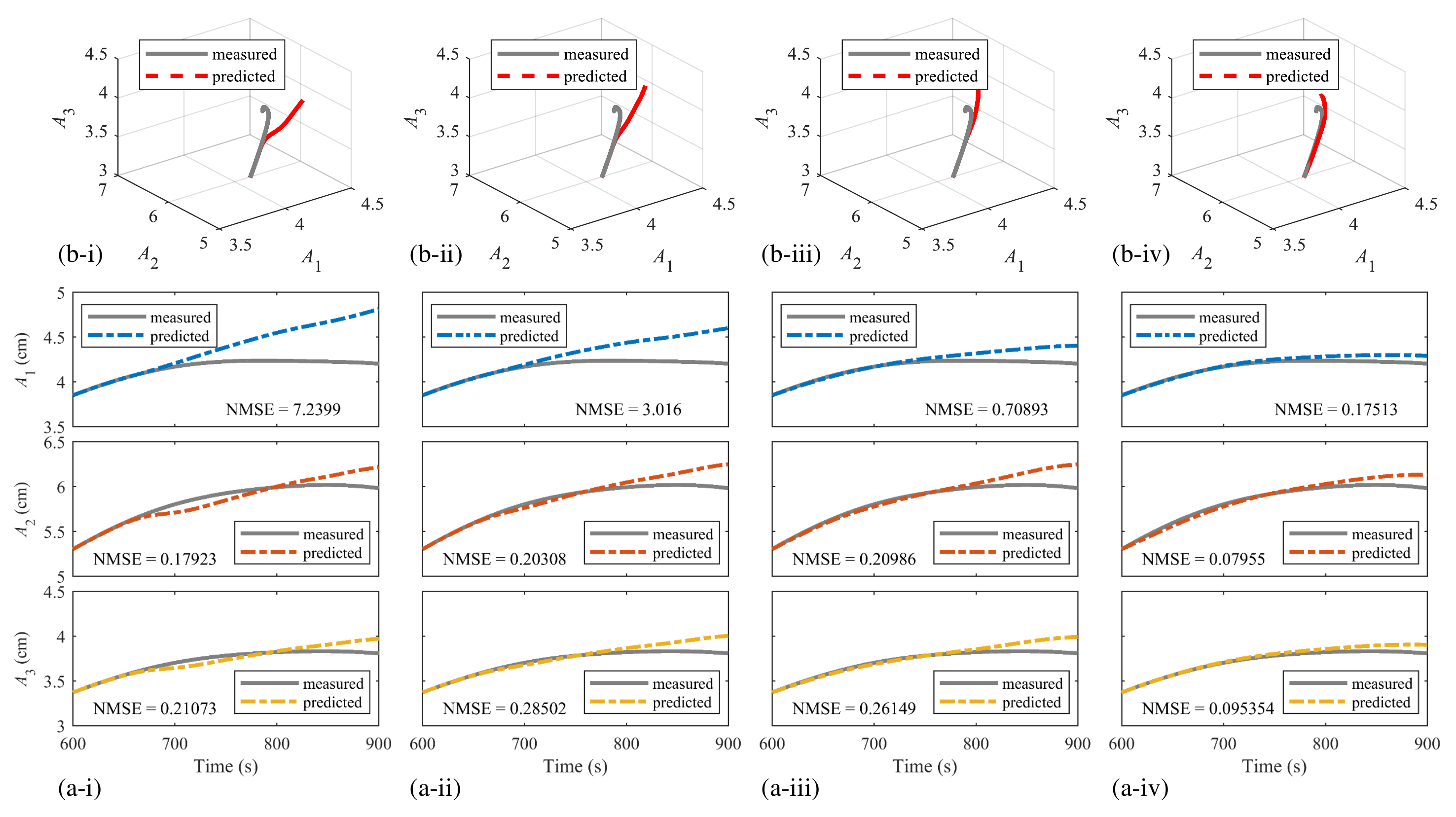}
	\caption{Forcasting for the range of similar dynamical regimes (600 s $\sim$ 900 s) in the VIV event No. 1 by solving the ODE with the time average of ${\bf{\Xi}}(t)$ over the first duration of different lengths of this range, with only the initial measured state ${\bf A}(t=600)$ and the measured wind history ${\bf U}(t)$ given. (a) Time histories of spatial components of ${\bf A}$. (b) Trajectory of ${\bf A}$ in the phase space. It should be noted that the used parameter ${\bf{\Xi}}$ for solving the ODE is constant rahter than time-dependent in the forecasting: (i) instantaneous ${\bf{\Xi}}(t)$ at 600 s, (ii) time average over 600 s $\sim$ 650 s, (iii) time average over 600 s $\sim$ 750 s, and (iv) time average over 600 s $\sim$ 900 s.}
	\label{fig:forecasting}
\end{figure*}

Forecasting future states of the bridge dynamics is critical for engineers to evaluate the safety of the bridge. Accordingly, we use the obtained model coefficients at only one moment ${\bf \Xi}(t_0)$, or time average of model coefficients over a short duration instead of an entire time-dependent parameter, to numerically solve the SINDy model.  This gives a forecast of the future state for the system after time $t_0$. The forecast expected to provide a short-time future state prediction while the model remains in the same dynamical regime as the initial state at $t_0$, or for the short duration of the time average of the the model coefficients ${\bf \Xi}$.  Eventually, the future state transitions to a different dynamical regime from the one used in the forecast, thus requiring an update of the forecast.   The range of the VIV event No. 1 (600 s $\sim$ 900 s), which is characterized by a strong self-excited effect and a large vibration amplitude, is used to evaluate the SINDy model forecasting, as shown in Fig.~\ref{fig:forecasting}. Using the model coefficients at time 600 s, we produce a good forecast to the future until approximately 700 s (see Fig.\ref{fig:forecasting}(a-i)). With an increasing length of time duration, one can average over the model parameters, thus increasing the overall forecasting accuracy.  This is indicated by a decreasing Normalized Mean Square Error (NMSE), as shown from Fig.~\ref{fig:forecasting}(a-i) to Fig.~\ref{fig:forecasting}(a-iv).  

These results show that the SINDy models are capable of producing accurate, short-time future state predictions of the system, allowing for enhanced monitoring of bridge dynamics.  Such tools can serve as critical assessment algorithms for real-time bridge monitoring.

\section{Discussion and Outlook}

In the present work, we have developed a data-driven model discovery technique that capitalizes on time series recordings used for bridge monitoring.   Specifically, we discover time-varying dynamical models of the nonlinear aerodynamics of a long-span suspension bridge from sparse, noisy sensor measurements which monitor the bridge at 1/4, 1/2 and 3/4 span.  Using the {\em sparse identification of nonlinear dynamics} (SINDy) algorithm, we are able to identify parsimonious, time-varying dynamical systems which result from vortex induced vibration (VIV) events in the bridge. Thus we are able to posit new, data-driven models highlighting the nonlinear fluid-structure interactions of the bridge structure with VIV events.  
These models extend the current state-of-the-art theory that is based upon weakly nonlinear dynamics.
The bridge dynamics is shown to have distinct, time-dependent modes of behavior, thus requiring parametric models to account for the diversity of dynamics.  Our method generates hitherto unknown bridge-wind interaction models that evolve in time and improve upon current theoretical and computational descriptions.  Our proposed method for real-time monitoring and model discovery allows us to move our model predictions beyond lab theory to practical engineering design.  The data-driven engineering designs can also be used to assess adverse engineering configurations that are susceptible to deleterious bridge-wind interactions.  With the rise of real-time sensor networks on major bridges, our model discovery methods can enhance an engineers ability to assess the nonlinear aeroelastic interactions of the bridge with its wind environment.

Due to the advent of networked sensors for real-time monitoring, the continuous assessment of modern bridge performance is now a reality.    Not only is it critical that bridges be monitored, e.g. for traffic monitoring and safety, but the rich time series recordings provided by the sensors allow bridge engineers to gain new understanding of the nonlinear aerodynamics of the bridge structure interactions with wind disturbances.  As such, new methods are now required in order to fully capitalize on these emerging {\em big data} applications.
Here, the discovery of nonlinear dynamical systems from time series recordings of a bridge has the potential to revolutionize engineering efforts and provide new theoretical insights that are beyond the scope of current, state-of-the-art bridge models.  We have shown that our proposed method can leverage (i) time-series measurements of a bridge sensor network, and (ii) the SINDy model discovery architecture to build data-driven models of the long-span suspension bridge. We find that the SINDy architecture is effective in identifying parsimonious, time-varying dynamical systems which result from VIV events in the bridge.  Thus we are able to posit new, data-driven models highlighting new aeroelastic interactions of the bridge structure with VIV events.

\section*{Acknowledgment}

Shanwu Li acknowledges funding support from the China Scholarship Council (Grant No. 201706120256).
Shujin Laima acknowledges funding support from the National Natural Science Foundation of China (Grant No. 51503138).
Hui Li acknowledges funding support from the National Natural Science Foundation of China (Grant No.51638007).
SLB and JNK acknowledge funding support from Defense Advanced Research Projects Agency (DARPA-PA-18-01-FP-125). 
EK also gratefully acknowledges support by the Washington Research Foundation, the Gordon and Betty Moore Foundation (Award \#2013-10-29) and the Alfred P. Sloan Foundation (Award \#3835).

\renewcommand{\bibfont}{\normalfont\footnotesize}

\bibliographystyle{plain}
\bibliography{myref.bib,references.bib}
\end{document}